\documentclass[twocolumn,floatfix,preprintnumbers,nofootinbib,superscriptaddress]{revtex4}

\usepackage{ulem}
\usepackage{bm}
\usepackage{times}
\usepackage{amssymb,  amsbsy,  amsmath,  amsfonts}
\usepackage{graphicx}
\usepackage{float}
\usepackage{color}
\usepackage{morefloats}
\usepackage{rotating}
\usepackage{srcltx}
\usepackage{slashed}
\usepackage{subfigure}
\usepackage{multirow}
\usepackage{verbatim}
\usepackage{hyperref}
\usepackage{tabularx}
\usepackage{braket}

\usepackage[outdir=./]{epstopdf}

\usepackage{graphicx}
\usepackage{amsmath}
\usepackage{amsfonts}
\usepackage{amssymb}
\usepackage{color}
\usepackage{multirow}
\usepackage{mathrsfs}
\usepackage{gensymb}

\usepackage{booktabs}  
\usepackage{threeparttable}

\usepackage{subfigure}

\newcommand\be{\begin{eqnarray}}
\newcommand\ee{\end{eqnarray}}

\begin{document}

\title{Roles of $\Delta(1232)$, $N^*(1520)$, and $N^*(1650)$ resonances in $\gamma p\to \pi^0 \pi^0 p$ reaction within an effective Lagrangian approach}

\author{Meng-Yuan Dai}
\affiliation{School of Physics, Zhengzhou University, Zhengzhou 450001, China}
\affiliation{Institute of Modern Physics, Chinese Academy of Sciences, Lanzhou 730000, China}

\author{Si-Wei Liu}
\affiliation{Institute of Modern Physics, Chinese Academy of Sciences, Lanzhou 730000, China}
\affiliation{School of Nuclear Sciences and Technology, University of Chinese Academy of Sciences, Beijing 101408, China}

\author{Cheng Chen}
\affiliation{Institute of Modern Physics, Chinese Academy of Sciences, Lanzhou 730000, China}
\affiliation{School of Nuclear Sciences and Technology, University of Chinese Academy of Sciences, Beijing 101408, China}

\author{De-Min Li}~\email{lidm@zzu.edu.cn}
\affiliation{School of Physics, Zhengzhou University, Zhengzhou 450001, China}

\author{En Wang}~\email{wangen@zzu.edu.cn}
\affiliation{School of Physics, Zhengzhou University, Zhengzhou 450001, China}

\author{Ju-Jun Xie}~\email{xiejujun@impcas.ac.cn}
\affiliation{Institute of Modern Physics, Chinese Academy of Sciences, Lanzhou 730000, China}
\affiliation{School of Nuclear Sciences and Technology, University of Chinese Academy of Sciences, Beijing 101408, China}
\affiliation{State Key Laboratory of Heavy Ion Science and Technology, Institute of Modern Physics, Chinese Academy of Sciences, Lanzhou 730000, China}
\affiliation{Southern Center for Nuclear-Science Theory (SCNT), Institute of Modern Physics, Chinese Academy of Sciences, Huizhou 516000, China}

\date{\today}

\begin{abstract}

Roles of the $\Delta (1232)$, ${N}^{*}(1520)$, and ${N}^{*}(1650)$ resonances in the $\gamma p\to{\pi }^{0}{\pi }^{0}p $ reaction near threshold is investigated within an effective Lagrangian approach. We have calculated the differential cross sections of the $\gamma p\to{\pi }^{0}{\pi }^{0}p$ reaction by including the contributions from the $\Delta (1232)$, ${N}^{*}(1520)$, and ${N}^{*}(1650)$ intermediate states decaying into $\pi^0 p$ via the $s$-channel nucleon pole and $t$-channel $\rho$ exchange, and found that the current experimental measurements can be well reproduced. The production of $\Delta(1232)$ is mainly from the mechanism of the $s$-channel nucleon pole, while the ${N}^{*}(1520)$ and ${N}^{*}(1650)$ are produced from the mechanism of the $t$-channel $\rho$ exchange. It is expected that more experimental data on the $\gamma p \to \pi^0 \pi^0 p$ reaction can be used to explore the properties of the low-lying excited baryon states and also the scalar $f_0(500)$ and $f_0(980)$ mesons.

\end{abstract}

\maketitle
\section{Introduction}

Studying the spectrum of the low-lying excited baryons and their decaying properties from the available experimental data is one of the most important topics in hadron physics~\cite{ParticleDataGroup:2024cfk,Klempt:2009pi,Crede:2013kia,Wang:2024jyk}. In the classical quark models, the baryon resonances are classified in shells according to the energy levels of the harmonic oscillator, and the quark model has achieved a great success in describing the baryon spectrum, especially for the ground states. There are many excited baryons predicted in the classical quark models~\cite{Isgur:1978xj,Capstick:1986ter,Capstick:1993kb,Loring:2001kx},  however only some of them were identified experimentally~\cite{ParticleDataGroup:2024cfk,Capstick:1992th,Shrestha:2012ep} and many predicted states have not been experimentally observed by now, which is the so-called `missing baryon problem'.
Meanwhile, for the low-lying excited baryons, the roper resonance $N^*(1440)$ ($J^P = 1/2^+$) belongs to the $N = 2$ shell, but it lies so much lower than the first orbitally excited nucleon states, such as $N^*(1535)$ ($J^P = 1/2^-$) and $N^*(1520)$ ($J^P = 3/2^-$) resonances~\cite{Zou:2009wp,Zhong:2024mnt}, which is the so-called `mass reversal problem'. Thus, on both the experimental and theoretical sides, much works remain to further establish the light-quark baryon spectrum and explore the nature of excited states.

For instance, the well-established isospin $I=3/2$ baryon $\Delta(1232)$ with spin-parity $J^P = 3/2^+$ mostly couples to the $\pi N$ channel, which implies it may have a large absolute value of the $\pi N$ compositeness~\cite{Sekihara:2015gvw}. For the $N^*(1520)$ resonance, it is the first orbitally excited (quark model prediction) nucleon resonance, and its coupling to the $\pi N$ channel is strong~\cite{ParticleDataGroup:2024cfk,An:2011sb,Zhong:2024mnt}. 
However, it is shown that the $N^*(1650)$ ($J^P = 1/2^-$) resonance gives an important contribution in the associate strangeness production reactions~\cite{COSY-TOF:2006tie,TOF:2010ygk,Li:2024rqb}. Thus, the strangeness component should be further considered for the structure of the excited baryons. 

It should be pointed out that, the photo-production processes provide a unique place to investigate these intermediate baryon resonances with strangeness component and small couplings to $\pi N$. For instance, it was shown that the intermediate nucleon and $\Delta$ excited states play a crucial role in the two-body reactions of $\gamma p \to p \pi^0$~\cite{Huang:2011as,CLAS:2019cpp,CLAS:2021cvy,Mai:2021aui}, $p \eta$~\cite{Zhong:2011ti,Suh:2018yiu,Mai:2021aui}, $p\eta'$~\cite{Huang:2009zzq,Zhong:2011ht,Huang:2012xj}, $p \omega$~\cite{CLAS:2009ffj,CLAS:2009hpc,Denisenko:2016ugz,Wei:2019imo}, $p \phi$~\cite{Kiswandhi:2010ub,Yang:2011es,Dey:2014tfa,Xie:2018kno,Kim:2021adl,Wu:2023ywu}, $K \Lambda$~\cite{delaPuente:2008bw,Anisovich:2014yza,Anisovich:2017ygb,Hunt:2018mrt,Kim:2018qfu}, $K \Sigma$~\cite{Mart:2019fau,Petrellis:2024ybj} $K^* \Lambda$~\cite{Oh:2006hm,Kim:2011rm,CLAS:2013qgi,Kim:2014hha,Wang:2017tpe,Wang:2019mid}, $K^* \Sigma$~\cite{Kim:2012pz,CLAS:2013qgi,Wang:2018vlv,Ben:2023uev}, $K \Lambda^*(1405)$~\cite{Niiyama:2008rt,Kim:2017nxg,Zhang:2021iez,Wang:2016dtb}, $K \Sigma^*(1385)$~\cite{Oh:2007jd,Niiyama:2008rt,Gao:2010hy,Chen:2013vxa,He:2013ksa,Wang:2015hfm,Wang:2020mdn}, and $K \Lambda^*(1520)$~\cite{LEPS:2009isz,Nam:2010au,Xie:2010yk,He:2012ud,Xie:2013mua,Xie:2013msa,Wang:2014jxb,He:2014gga,Wei:2021qnc}. Detailed investigations of these photo-production reactions have substantially enhanced our understanding of both the reaction mechanisms and the properties of intermediate baryon resonances.


Recently, the differential cross sections for the reaction $\gamma p \to \pi^0 \pi^0 p$ were firstly measured using a linearly polarized photon beam with energy from reaction threshold up to 2.4~GeV by the LEPS2/BGOegg Collaboration~\cite{LEPS2BGOegg:2023ssr}. Indeed,
 the two-pion photo-production processes off proton targets have been measured in numerous experiments~\cite{Braghieri:1994rf,Harter:1997jq,Wolf:2000qt,CLAS:2009ngd,CBELSATAPS:2015kka}. For the $\gamma p \to \pi^0 \pi^0 p$ reaction, it provides interesting details because many Born terms are strongly suppressed and most nucleon resonances as well as the $\rho^0(770)$ cannot directly decay into two neutral pions. Although the main purpose of Ref.~\cite{LEPS2BGOegg:2023ssr} is to investigate the nature of the $f_0(980)$ decaying into $\pi^0\pi^0$ from $\gamma p \to f_0(980) p \to \pi^0 \pi^0 p$ reaction,  the uncertainties of the experimental data on the invariant $\pi^0 \pi^0$ mass distributions are large and the extracted mass and width of $f_0(980)$ meson are lower than the averaged values quoted in the Reviwe of Particle Physics (RPP)~\cite{ParticleDataGroup:2024cfk}. The production of $f_0(980)$ meson in the $\gamma p \to f_0(980) p \to \pi^0 \pi^0 (K \bar{K}) p$ reactions was studied theoretically in Refs.~\cite{Ji:1997fb,Donnachie:2015jaa,Lee:2016vlw,Xing:2018axn,Wei:2024lne} within an effective Lagrangian approach. It is worthy to mention that the experimental measurements of Ref.~\cite{LEPS2BGOegg:2023ssr} show that the contributions of $\Delta(1232)$, $N^*(1520)$, and $N^*(1650)$ resonances to the final $\pi^0 p$ channel are important and there are clear bump structures for these resonances in the $\pi^0 p$ invariant mass distributions, which could be used to investigate the  roles of $\Delta(1232)$, $N^*(1520)$, and $N^*(1650)$ resonances in this process. 

In the present work, based on the new experimental measurements by the LEPS2/BGOegg Collaboration of Ref.~\cite{LEPS2BGOegg:2023ssr}, we study the roles of $\Delta(1232)$, $N^*(1520)$, and $N^*(1650)$ resonances in the $\gamma p \to \pi^0 \pi^0 p$ reaction within an effective Lagrangian approach, one important theoretical method for describing various processes in the resonances production region~\cite{Xie:2005sb,Xie:2007vs,Xie:2007qt,Xie:2008ts,Xie:2010md,Liu:2011sw,Liu:2012ge,Xie:2013db,Xie:2013wfa,Lu:2013jva,Lu:2014rla,Xie:2014kja,Xie:2014tra,Xie:2014zga,Wu:2014yca,Cao:2014mea,Wang:2014ofa,Lu:2014yba,Huang:2014gxa,Xiao:2015zja,Xie:2015zga,Garzon:2015zva,Xie:2015wja,Lu:2015fva,Huang:2016tcr,Cheng:2016hxi,Huang:2016ygf,Zhang:2017eui,Zhang:2018kdz,Zhou:2019eaf,Wang:2022vjm}. For the production of $\Delta (1232)$, ${N}^{*}(1520)$, and ${N}^{*}(1650)$ resonances, both $s$-channel proton pole and $t$-channel $\rho^0$ exchange processes are taken into account. It is shown that the new experimental measurements on the $\pi^0 p$ invariant mass distributions of Ref.~\cite{LEPS2BGOegg:2023ssr} can be well reproduced with the contributions of $\Delta(1232)$ in the $s$-channel process and $N^*(1520)$ and $N^*(1650)$ in the $t$-channel process. In this respect, we show in this work how the new measurements about $\gamma p \to \pi^0 \pi^0 p$ reaction could be used to extract the properties of these baryon resonances decaying into $\pi^0 p$. Note that we will leave the study of scalar meson $f_0(980)$ in $\gamma p \to \pi^0 \pi^0 p$ reaction to further work when more precise experimental data are available.

This article is organised as follows. We will give the theoretical formalism of this work in Sect.~\ref{sec:form}, then the numerical results and discussions are given in Sect.~\label{sec:result}. A short summary is given in the last section. 

\section{Theoretical formalism}
\label{sec:form}
In this section, we introduce the theoretical formalism to calculate the differential scattering cross section for the $\gamma p\to {\pi }^{0}R\to{\pi }^{0}{\pi }^{0}p $ [$R\equiv\Delta (1232)$, ${N}^{*}(1520)$, or ${N}^{*}(1650)$] reaction within the effective Lagrangian approach, which is widely used to investigate the scattering reactions in the resonances production region~\cite{Xie:2013msa,Wang:2014jxb,He:2014gga,Wei:2021qnc,Wang:2017hug,Zhao:2019syt}.

 \begin{figure}[htbp]
\centering
\subfigure[]{\includegraphics[scale=0.45]{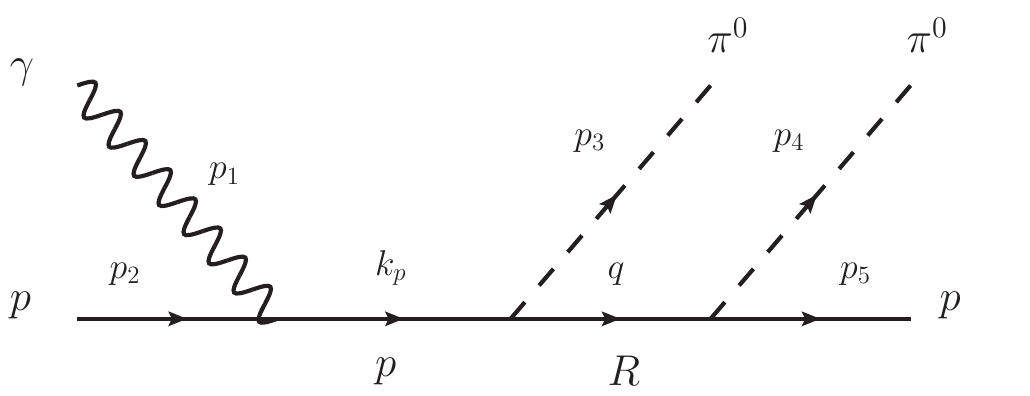}}
\subfigure[]{\includegraphics[scale=0.45]{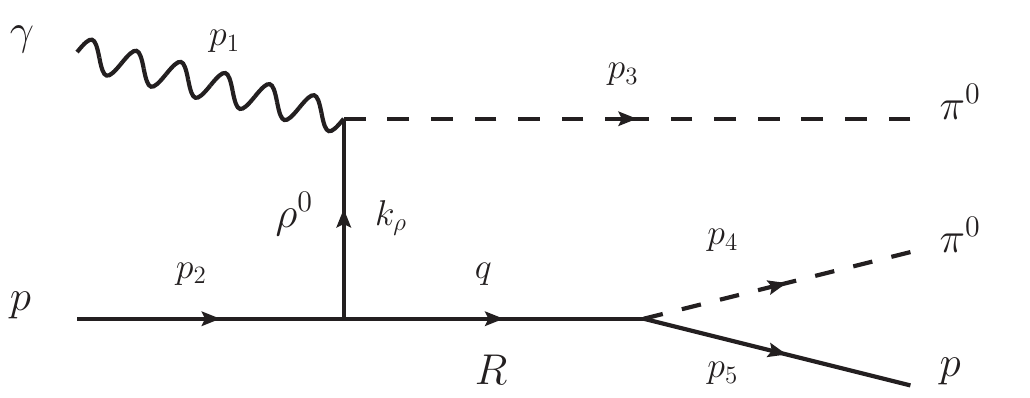}}
\caption{Feynman diagrams for the reaction $\gamma p\to {\pi }^{0}R \to {\pi }^{0}{\pi }^{0}p$ [$R\equiv\Delta (1232)$, ${N}^{*}(1520)$, or ${N}^{*}(1650)$]. It consists of $s$-channel nuclepn pole (a) and $t$-channel $\rho^0$ meson exchange (b). We also show the definition of the kinematical ($p_1$, $p_2$, $p_3$, $p_4$, $p_5$) used in the present calculation. In addition, we use $k_p = p_1 + p_2$, $k_\rho = p_1 - p_3$, and $q = p_4 + p_5$.} \label{fig:feyndigrams}
\end{figure}

The basic tree level Feynman for the $\gamma p \to \pi^0 \pi^0 p$ reaction are presented in Fig.~\ref{fig:feyndigrams}, which includes the $s$-channel proton pole [Fig.~\ref{fig:feyndigrams}(a)] and $t$-channel $\rho^0$ exchange process [Fig.~\ref{fig:feyndigrams}  (b)]. To compute the contributions of those terms shown in Fig.~\ref{fig:feyndigrams}, we use the following effective interaction Lagrangian densities for the $\gamma p p$ and $\gamma \rho \pi$ vertices as usded in Refs.~\cite{Tarasov:2013yma,Oh:2003aw,Oh:2000zi},
 \begin{eqnarray}
{\mathcal{L}}_{\gamma pp}&=&- e\overline{p}\left[ \slashed {A}-\frac{{\mathcal{K}}_{p}}{2{m}_{N}}{\sigma }^{\alpha\beta }({\partial }_{\beta}{A}_{\alpha})\right]p, \\
{\mathcal{L}}_{\rho \gamma \pi }&=&\frac{e{g}_{\rho \gamma \pi }}{{m}_{\rho }}{\varepsilon }^{\mu \nu \alpha \beta }{\partial }_{\mu}{\rho}_{\nu}{\partial }_{\alpha}{A}_{\beta}\pi ,  
\end{eqnarray}
where $e=\sqrt{4\pi \alpha }$ with $\alpha = 1/137.036$ the electromagnetic fine structure constant. ${A}_{\beta}$, $p$, $\bar{p}$, $\pi $, and ${\rho}_{\nu}$ denote the fields of the photon, proton, antiproton, $\pi$, and $\rho$, respectively. The magnetic moment ${{\mathcal{K}}_{p}}=1.5$ which is taken from Ref.~\cite{Xing:2018axn}. The coupling constant ${g}_{\rho \gamma \pi }$ can be obtained from the partial decay width,
\begin{eqnarray}
{\varGamma }_{{\rho}^{0} \rightarrow {\pi }^{0}\gamma} = \frac{e^2 g^2_{\rho \gamma \pi}}{96\pi} \frac{(m^2_{\rho^0} - m^2_{\pi^0})^3}{m^5_{\rho^0}}.
\end{eqnarray}
With $m_{\rho^0} = 775.26$~MeV, $m_{\pi^0} = 134.98$~MeV, and $\Gamma_{\rho^0 \to \pi^0 \gamma} = 70.08$~keV~\cite{ParticleDataGroup:2024cfk,Zhang:2018kdz}, one can easily get ${g}_{\rho \gamma \pi }$=0.57. Since we do not consider the interference terms between different resonances, the theoretical results will not depend on the sign of the coupling constant ${g}_{\rho \gamma \pi }$.

\begin{table}[htbp]
    \centering
    \caption{The coupling constants $g_{\pi NR}$ used in this work.}
    \begin{tabular}{c c  c   c     c} \hline \hline
        State &\parbox{1.5cm} {Mass(MeV)}&\parbox{2cm}{Width (MeV)}  & \parbox{1cm}{$\mathcal{B}$} & \parbox{1cm}{$g^2_{\pi N R}$ }\\\hline
        $\Delta (1232)$ &1210 & 100  & 0.994 & 1.59 \\
        ${N}^{*}(1520)$ &1505  & 110  & 0.60 & 2.55 \\
        ${N}^{*}(1650)$ &1670 & 110  & 0.60 & 0.41 \\    
        \hline \hline
    \end{tabular}
    \label{tab:pi_NR_coupling}
\end{table}

For the $\pi NR$ vertices, we use the effective interaction Lagrangian commonly adopted in Refs.~\cite{Zou:2002yy,Lu:2014yba,Lu:2015pva}:
\begin{align}
{\mathcal{L}}_{\pi N}^{\Delta (1232)}&= - \frac{{g}_{\pi N\Delta}}{{m}_{\pi }} \bar{{R}}_{\mu }\vec{\tau }\cdotp {\partial }^{\mu }\vec{\pi }N + h.c.,\\
{\mathcal{L}}_{\pi N}^{{N}^{\ast }(1520)}&= - \frac{{g}_{\pi N N^*(1520)}}{{m}_{\pi }} \bar{{R}}_{\mu }{\gamma }_{5}\vec{\tau }\cdotp {\partial }^{\mu }\vec{\pi }N + h.c.,\\
{\mathcal{L}}_{\pi N}^{{N}^{\ast }(1650)}&= i{g}_{\pi N N^*(1650)} \bar{{R}}\vec{\tau }\cdotp \vec{\pi }N + h.c.,
\end{align}
where $\vec{\tau}$ is the Pauli matrix and $\vec{\pi} = (\pi_1,\pi_2,\pi_3)$. We take $\pi^+ = (\pi_1 - i \pi_2)/\sqrt{2}$, $\pi^- = (\pi_1 + i \pi_2)/\sqrt{2}$, and $\pi^0 = \pi_3$. In addition, the coupling constants $g_{\pi NR}$ are obtained from the following partial decay widths,
\begin{align}
\varGamma_{\Delta(1232) \to N{\pi }}&=\frac{{g}_{\pi N \Delta(1232)}^{2}}{4\pi}\frac{({E}_{N}+{m}_{N})}{{m}_{\Delta(1232)}{m}_{\pi }^{2}}{\left | {\vec{p}}_{N}\right | }^{3}, \\
\varGamma_{N^*(1520) \to N{\pi } } &=\frac{{g}_{\pi N N^*(1520)}^{2}}{4\pi}\frac{({E}_{N}-{m}_{N})}{{m}_{N^*(1520)}{m}_{\pi }^{2}}{\left | {\vec{p}}_{N}\right | }^{3}, \\
\varGamma_{N^*(1650) \to N{\pi} } &=\frac{3{g}_{\pi N N^*(1650)}^{2}}{4\pi}\frac{({E}_{N}+{m}_{N})}{{m}_{N^*(1650)}}\left | {\vec{p}}_{N}\right |,
\end{align}
with
\begin{eqnarray}
\left | {\vec{p}}_{N}\right |=\frac{{\lambda }^{\frac{1}{2}}({m}_{R}^{2}, {m}_{\pi}^{2}, {m}_{N}^{2})}{2 {m}_{R}}, \\
 E_{N}=\frac{{m}_{R}^{2}+ {m}_{N}^{2}-{m}_{\pi}^{2}}{2 {m}_{R}},
\end{eqnarray}
where the  K$\ddot{a}$ll$\acute{e}$n function is $\lambda (a, b, c)={a}^{2}+{b}^{2}+{c}^{2}-2ab-2ac-2bc$, and $ {m}_{R}$ represents the masses of the $\Delta(1232)$, $N^*(1520)$, or $N^*(1650)$ resonance. 

With masses and partial decay widths of $\Delta(1232)$, $N^*(1520)$, and $N^*(1650)$ resonances to the $\pi N$ channel~\cite{ParticleDataGroup:2024cfk}, the coupling constants $g^2_{\pi NR}$ can be obtained as listed in Table~\ref{tab:pi_NR_coupling}.

Next, for the $\rho N R$ vertices, we use the effective interaction Lagrangian densities commonly adopted in previous works~\cite{Lu:2015pva,Zou:2002yy,Matsuyama:2006rp},
\begin{eqnarray}
{\mathcal{L}}_{\rho N}^{\Delta (1232)}  &=&   \frac{-i{g}_{\rho N\Delta}}{{m}_{\rho}} \bar{N}{\gamma }^{\sigma}{\gamma }_{5}\vec{\tau }\cdotp [\partial _{\mu } \vec{\rho }_{\sigma}-\partial _{\sigma } \vec{\rho }_{\mu} ]{R}^{\mu }+ h.c., \nonumber \\
&&  \\
{\mathcal{L}}_{\rho N}^{{N}^{\ast }(1520)}&=& {g}_{\rho NN^*(1520)}\bar{N}\vec{\tau }\cdotp \vec{\rho }^{\,\mu }{R}_{\mu } + h.c., \\
{\mathcal{L}}_{\rho N}^{{N}^{\ast }(1650)}&=&i{g}_{\rho NN^*(1650)}\bar{{N}}{\gamma }_{5}\left({\gamma }_{\mu } - \frac{{q}_{\mu }\slashed {q}}{q^2}\right)\vec{\tau }\cdotp \vec{\rho }^{\,\mu }R + h.c.. \nonumber \\
\end{eqnarray}
Since the mass threshold of $\rho N$ is higher than the masses of $\Delta(1232)$, $N^*(1520)$, and $N^*(1650)$ resonances, the coupling constants $g_{\rho NR }$ can not extracted from the decay widths, and will be determined with the current experimental data in following. 

With the ingredients presented above, the total scattering amplitudes of $\gamma p\to {\pi }^{0}R\to{\pi }^{0}{\pi }^{0}p $ reaction in the $s$-channel and $t$-channel can be written as

\begin{widetext}
\begin{eqnarray}
{\mathcal{M}}_{s}^{\Delta (1232)}&=& ie\frac{{g}_{\pi N \Delta}^{2}}{m_{\pi }^{2}}{F}_{p}(k_p^2){F}_{\Delta}({q^2})\bar{u}({p}_{5}){G}_{s=\frac{3}{2}}^{\omega \sigma }({q}){G}_{s=\frac{1}{2}}({k_p}) \left({\gamma }^{\mu }-{\varGamma }_{ c}^{\mu }-\frac{{\mathcal{K}}_{p}}{2{m}_{N}}{\gamma }^{\mu }\slashed{p}_{1}\right)u({p}_{2}){\varepsilon }_{\mu }({p}_{1}){p}_{3\omega }{p}_{4\sigma }\nonumber\\
  && + (\text{exchange terms with}~ {p}_{3}\leftrightarrow {p}_{4}),  \\
{\mathcal{M}}_{s}^{{N}^{*}(1520)}&=& ie\frac{{g}_{\pi NN^*(1520)}^{2}}{m_{\pi }^{2}}{F}_{p}(k_p^2){F}_{N^*(1520)}({q^2})\bar{u}({p}_{5}){\gamma }_{5}{G}_{s=\frac{3}{2}}^{\omega \sigma }({q}) {\gamma }_{5}{G}_{s=\frac{1}{2}}({k_p})\left({\gamma }^{\mu }-{\varGamma }_{ c}^{\mu }-\frac{{\mathcal{K}}_{p}}{2{m}_{N}}{\gamma }^{\mu }\slashed {p}_{1}\right)\nonumber\\
  &&u({p}_{2}){\varepsilon }_{\mu }({p}_{1}){p}_{3\omega  }{p}_{4\sigma }  + (\text{exchange terms with}~ {p}_{3}\leftrightarrow {p}_{4}),  \\
{\mathcal{M}}_{s}^{{N}^{*}(1650)}&=& ie{g}_{\pi NN^*(1650)}^{2}{F}_{p}(k_p^2){F}_{N^*(1650)}({q^2})\bar{u}({p}_{5}){G}_{s=\frac{1}{2}}({q}){G}_{s=\frac{1}{2}}({k_p})\left({\gamma }^{\mu  }-{\varGamma }_{ c}^{\mu }-\frac{{\mathcal{K}}_{p}}{2{m}_{N}}{\gamma }^{\mu }\slashed {p}_{1}\right)u({p}_{2})\nonumber\\
  &&{\varepsilon }_{\mu }({p}_{1}) + \left(\text{exchange terms with}~ {p}_{3}\leftrightarrow {p}_{4}\right),  \\
{\mathcal{M}}_{t}^{{N}^{*}(1520)}&=&\frac{ie{g}_{\pi NN^*(1520)}{g}_{\rho NN^*(1520)}{g}_{\rho\gamma\pi}}{{m}_{\pi }{m}_{\rho }}F(k_\rho^2)F_{N^*(1520)}({q}^2)\bar{u}({p}_{5}){\gamma }_{5}{p}_{4\sigma} {G}_{s=\frac{3}{2}}^{\sigma\omega }({q}){{G}_{s=1}}_{\omega \nu}(k_\rho)k_{\rho\mu}p_{1\alpha}\varepsilon^{\mu \nu \alpha\beta}\nonumber\\
  &&\varepsilon _{\beta}(p_{1}){u}({p}_{2}) + (\text{exchange terms with}~ {p}_{3}\leftrightarrow {p}_{4}), \\
{\mathcal{M}}_{t}^{{N}^{*}(1650)}&=&\frac{e{g}_{\pi NN^*(1650)}{g}_{\rho NN^*(1650)}{g}_{\rho\gamma\pi}}{{m}_{\rho }}F(k_\rho^2)F_{N^*(1650)}({q}^2)\bar{u}({p}_{5}){G}_{s=\frac{1}{2}}({q}){\gamma}_{5}\left(\gamma^{\sigma}-\frac{{q}^{\sigma}\slashed {q}}{q^2}\right){{G}_{s=1}}_{\sigma\nu}(k_\rho)k_{\rho\mu}p_{1\alpha}\nonumber\\
  &&\varepsilon ^{\mu \nu \alpha\beta}\varepsilon _{\beta}(p_{1}){u}({p}_{2}) + (\text{exchange terms with}~ {p}_{3}\leftrightarrow {p}_{4}) ,
\end{eqnarray}
\end{widetext}
where the propagator  ${G}_{s=1}$ for the $\rho$ meson with spin $s=1$ is given by~\cite{Xu:2024xso},
\begin{eqnarray}
{G}_{s=1}^{\mu \nu }(k_\rho)=i\frac{-{g}^{\mu\nu} + {k^\mu_\rho k^\nu_\rho }/{k_\rho^2} }{k_\rho^2-m^2_\rho},
\end{eqnarray}
the propagator $G_{s=\frac{1}{2}}$ is given by~\cite{Tsushima:1998jz,Lu:2014rla},
\begin{eqnarray}
{G}_{s=\frac{1}{2}}({q})=\frac{i(\slashed {{q}}+{{m}_{R}})}{q^2 - {{{m}^2_{R}}} +i{{m}_{R}}{{\mathrm \varGamma }_{R}}},
\end{eqnarray}
for the nucleon pole and $N^*(1650)$ resonance, and the propagator ${G}_{s=\frac{3}{2}}^{\omega\sigma }$ is given by Refs.~\cite{Tsushima:1998jz, Penner:2002ma}:
\begin{eqnarray}
{G}_{s=\frac{3}{2}}^{\omega\sigma }({q})=\frac{i(\slashed {{q}}+{{m}_{R}}){P}^{\omega\sigma}({q})}{{q}^{2}-{{{m}^2_{R}}} + i{{{m}_{R}}}{\varGamma }_{R}},  
\end{eqnarray}
for the $\Delta(1232)$ and $N^*(1520)$ resonances, where
\begin{eqnarray}
{P}^{\omega\sigma }({q})&=&-{g}^{\omega\sigma }+\frac{1}{3}{\gamma }^{\omega }{\gamma }^{\sigma }+\frac{1}{3{{m}_{R}}}({\gamma }^{\omega }{{q}}^{\sigma }-{\gamma }^{\sigma }{{q}}^{\omega}) \nonumber\\
  && + \frac{2}{3{{m}_{R}^{2}}}{{q}}^{\omega }{{q}}^{\sigma }.
\end{eqnarray}
In these above equations, ${{m}_{R}}$ and ${\mathrm{\varGamma }_{R}}$ represent the masses and total widths of the intermediate baryon states.

Note that the contact term involving $\varGamma^\mu_c$ is taken into account to keep the scattering amplitudes gauge invariant. By including the following term $\varGamma^\mu_c$ in the scattering amplitudes ${\mathcal{M}}_{s}$~\cite{Haberzettl:2006bn},
\begin{eqnarray}
{\varGamma }_{ c}^{\mu}=\frac{\slashed {p}_{1}{p}_{2}^{\mu}}{{p}_{1} \cdot {p}_{2}},
\end{eqnarray}
it is easy to show that the total amplitude ${\cal M}_{\rm total}$ satisfies the gauge invariance
\begin{eqnarray}
p_1 \cdot {\cal M}_{\rm total} = 0,
\end{eqnarray}
with
\begin{eqnarray}
{\cal M}_{\rm total} &=& {\mathcal{M}}_{s}^{\Delta (1232)} +  {\mathcal{M}}_{s}^{{N}^{*}(1520)} + {\mathcal{M}}_{s}^{{N}^{*}(1650)} \nonumber \\
&& + {\mathcal{M}}_{t}^{{N}^{*}(1520)} + {\mathcal{M}}_{t}^{{N}^{*}(1650)}.
\end{eqnarray}
And it is found that the contribution of $\Delta(1232)$ in the $t$-channel is zero, because the $\Delta N \rho$ coupling vanishes when used in connection with the $\rho \pi \gamma$ vertex to avoid the gauge invariance problem.

Furthermore, because hadrons are not point-like particles, the form factors of hadrons must be taken into account. The relevant off-shell form factor is also used for the exchanged particles to take into account the internal structure of hadrons and off-shell effects~\cite{Xing:2018axn,Shklyar:2005xg,Feuster:1997pq}. For the $\rho$ meson exchange, we introduce the form factor as in Refs.~\cite{Brockmann:1990cn,Machleidt:1989tm,Machleidt:1987hj}:
\begin{eqnarray}
F(k_\rho^2)=\frac{\Lambda_\rho^2-m_\rho^2}{\Lambda_\rho^2-k_\rho^2},
\end{eqnarray}
where $\Lambda_\rho$, $m_\rho$ and $k_\rho$ are the cutoff parameter, mass and four-momentum of the exchanged $\rho$ meson, respectively. In this work, we take $\Lambda_\rho$ =1.3 GeV~\cite{Xie:2007vs}. 

For the intermediate baryon resonances, we adopt the form factor as used in Refs.~\cite{Feuster:1997pq,Penner:2002ma,Wei:2024lne,Wang:2017tpe},
\begin{eqnarray}
{F_{R}}({q^2})=\frac{{\Lambda^{4}}}{{\Lambda^{4}}+({{{q}}^{2}-{m^{2}_R}})^{2}},  
\label{form factor}
\end{eqnarray}
where $\Lambda$, $m_R$ and $q$  are the cutoff parameter, mass and four-momentum of the exchanged baryon, respectively. In this calculation, the free cutoff parameters are adopted as follows: $\mathrm {{\Lambda }}=1.0$~GeV for $\Delta(1232)$, $\mathrm {{\Lambda }}=2.0 $~GeV for $N^*(1520)$ and $N^*(1650)$ resonances~\cite{Xie:2007qt,Lu:2015pva}, and $\mathrm {{\Lambda }} = 1.1 $~GeV for the proton pole~\cite{Xing:2018axn}. 

\section{Numerical results and discussions}

\begin{figure*}[htbp]
\centering
\includegraphics[scale=0.7]{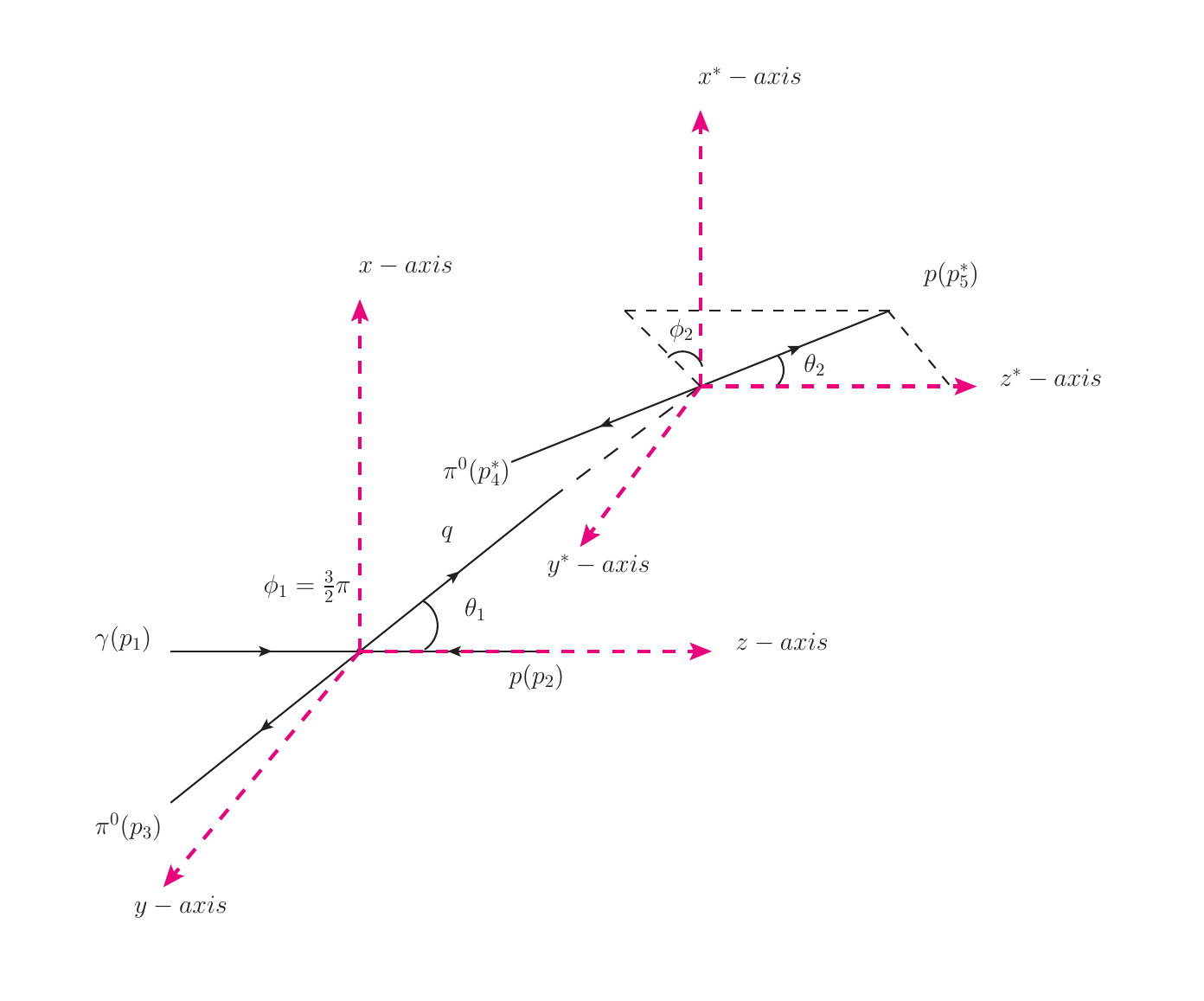} 
\caption{Definitions of the variables in the phase space integration of the $\gamma p\to{\pi }^{0}{\pi }^{0}p$ reaction.} \label{fig:phase-angle}
\end{figure*}

In this section,  we will show the numerical results for the $\gamma p\to {\pi }^{0}R\to{\pi }^{0}{\pi }^{0}p $ reaction. Considering the three-body phase space of the $\gamma p\to{\pi }^{0}{\pi }^{0}p $ reaction (see more details in Fig.~\ref{fig:phase-angle}), one can get~\cite{ParticleDataGroup:2024cfk}
 \begin{eqnarray}
\frac{{d}\sigma }{d{m}_{{\pi }^{0}p}} (W) &=& \frac{\left | \vec{p}_{3}\right |\left | \vec{{p}}^{\,*}_{4}\right |}{{2}^{10}{\pi }^{5} W (W^2-m^2_p)} \int d{\Omega}_{1}\int d{\Omega}^*_{2} \nonumber \\
&& \times \frac{1}{4}\sum_{\rm spins}{{\left | {\mathcal{M}}_{\rm total}\right |^{2}}},   \label{dcs}
\end{eqnarray}
where $W$ is the invariant mass of the $\gamma p$ system. $\Omega_{1}$ and $\Omega^*_{2}$ are the solid angles in the center-of-mass system of the $\gamma p$ collision and in the center-of-mass system of the two-body $\pi^0 p$ final state, respectively. The $\vec{p}_{3}$ and $\vec{{p}}^{\,*}_{4}$ are the three-momenta of first $\pi^{0}$ (from the electromagnetic vertex) and second $\pi^0$ (from baryon resonance decay) in the center-of-mass system of $\gamma p$ collision and the center-of-mass system of the two-body $\pi^0 p$ final state, respectively, which are given by
\begin{eqnarray}
|\vec{p}_3| &=& \frac{{\lambda }^{\frac{1}{2}}(W^{2}, {m}_{{\pi }^{0}}^{2}, {m}_{\pi^0 p}^{2})}{2 W},  \\
|\vec{p}^{\,*}_4| &=& \frac{{\lambda }^{\frac{1}{2}}({m}_{\pi^0 p}^{2}, {m}_{{\pi }^{0}}^{2}, {m}_{p}^{2})}{2 {m}_{\pi^0 p}} .
\end{eqnarray}

In order to compare the theoretical calculations with the experimental measurements from the LEPS2/BGOegg Collaboration in Ref.~\cite{LEPS2BGOegg:2023ssr}, we calculate the differential cross section $d\sigma /d{m}_{{\pi }^{0}p}$ by 
\begin{eqnarray}
\frac{d\sigma }{d{m}_{{\pi }^{0}p}} = \frac{\int_{{W}_\mathrm {min}}^{{W}_\mathrm {max}}\frac{{d}\sigma }{d{m}_{{\pi }^{0}p}}(W) dW }{W_\mathrm {max} - W_\mathrm {min}},
\end{eqnarray}
with $W_\mathrm {min}$ =1898~MeV and $W_\mathrm {max}$ = 2320~MeV, which are the total energy regions of the experimental measurements given in Ref.~\cite{LEPS2BGOegg:2023ssr}. One the other hand, a constant background term is also included, thus, the $\left | {\mathcal{M}}_{\rm total}\right |^{2}$ is written as
 \begin{eqnarray}
{\left | {\mathcal{M}}_{\rm total}\right |^{2}} &=& c_1 \left ( {\left | {\mathcal{M}}_{s}^{\Delta (1232)}\right |}^{2}
+{\left | {\mathcal{M}}_{s}^{{N}^{*}(1520)}\right |}^{2} \right. \nonumber\\
&& \left. +{\left | {\mathcal{M}}_{s}^{{N}^{*}(1650)}\right |}^{2} +{\left | {\mathcal{M}}_{t}^{{N}^{*}(1520)}\right |}^{2} \right. \nonumber\\
&& \left.  + {\left | {\mathcal{M}}_{t}^{{N}^{*}(1650)}\right |}^{2} + c_2 \right ),
\label{the total decay amplitude}
\end{eqnarray}
where the interference terms between different resonances are ignored. Besides, it is found that the contributions of the ${N}^{*}(1520)$ and ${N}^{*}(1650)$ resonances in the $s$-channel are rather small and can be also ignored. The factors $c_1$ and $c_2$ are introduced to scale the theoretical differential cross sections to match the experimental measurements of the signal yields.

\begin{figure}[htbp]
    \centering
    \includegraphics[scale=0.37]{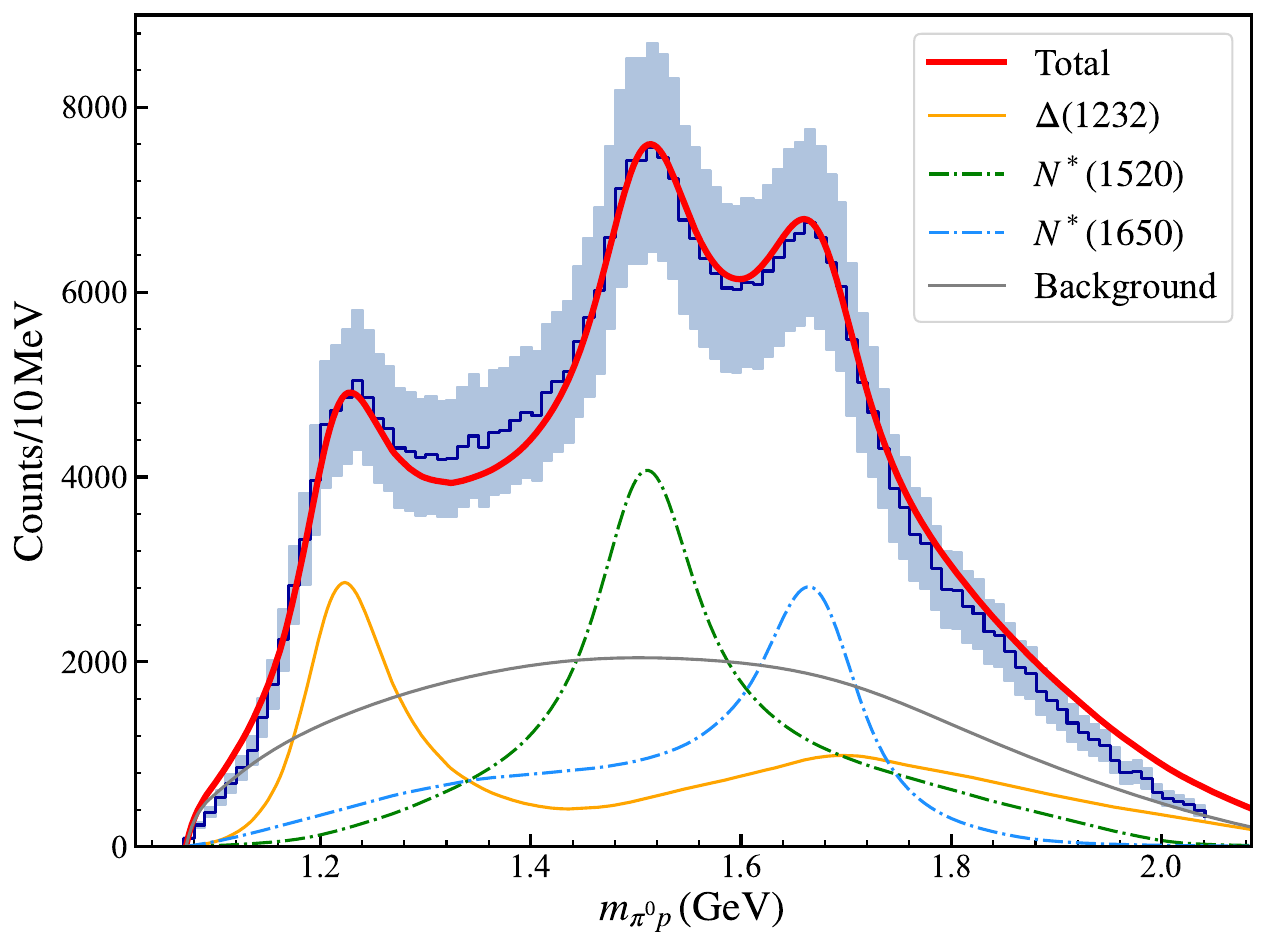}
\caption{Invariant mass distribution of ${\pi }^{0} p$ in the energy range $ 1898< W < 2320 $ MeV compared with the experimental measurements taken from Ref.~\cite{LEPS2BGOegg:2023ssr}.}
    \label{fig:massa-1989-2320-pi0p}
\end{figure}

We show the theoretical results for the $\pi^0 p$ invariant mass distributions  at total energy region of $ 1898< W < 2320 $~MeV in Fig.~\ref{fig:massa-1989-2320-pi0p}, where one can see that the experimental data can be well reproduced. The experimental values are represented in blue, while the light blue error band reflects the 15\% uncertainty of the experimental data~\cite{LEPS2BGOegg:2023ssr}. And it is clear seen that the three prominent peaks correspond to the $\Delta (1232)$, ${N}^{*}(1520)$, and ${N}^{*}(1650)$ resonances. The theoretical numerical results are obtained with $c_1 = 1.08\times 10^6$, $c_2 = 5.14 \times 10^{2}$, and the coupling constants $g_{{\rho}NN^*(1520)}= 33.34$ and $g_{{\rho}NN^*(1650)} = 30.37$.

Next we turn to the invariant mass distributions of the $\pi^0 \pi^0$ of this process\footnote{Since we donot consider the contributions from $f_0(500)$ and $f_0(980)$ mesons, the theoretical results of $\pi^0\pi^0$ invariant mass distributions can be viewed as the background contributions.}. It is worth to mention that, on the experimental side, to improve the signal-to-noise ratio of the $f_0(980)$ meson, the LEPS2/BGOegg Collaboration screened the events, ultimately leading to the number of events in the sample being 133,000, which means that approximately 27$\% $ of the event numbers were lost. In the theoretical calculations for the $\pi^0\pi^0$ invariant mass spectrum, an extra global factor 73$\% $ was taken into account in order to more effectively compare our theoretical results with the number of experimental events. 
Additionally, based on the experimental data, the area ratio of the high-energy region ($2110< W < 2320 $~MeV) to the low-energy region  ($ 1898< W< 2110 $~MeV) in the spectrum is 0.75 (see more details in Ref.~\cite{LEPS2BGOegg:2023ssr}), which is,
\begin{eqnarray}
&&0.73\sum_{m_{\pi^0p}}\mathrm{Events}= \sum_{m_{\pi^0\pi^0}^{\rm low}}\mathrm{Events}+\sum_{m_{\pi^0\pi^0}^{\rm high}}\mathrm{Events},\nonumber\\
&&\frac{\sum_{m_{\pi^0\pi^0}^{\rm low}}\mathrm{Events}}{\sum_{m_{\pi^0\pi^0}^{\rm high}}\mathrm{Events}}=0.75 .
\label{Exp-the area ratio of the higher and lower}
\end{eqnarray}

\begin{figure}[htbp]
    \centering
    \includegraphics[scale=0.37]{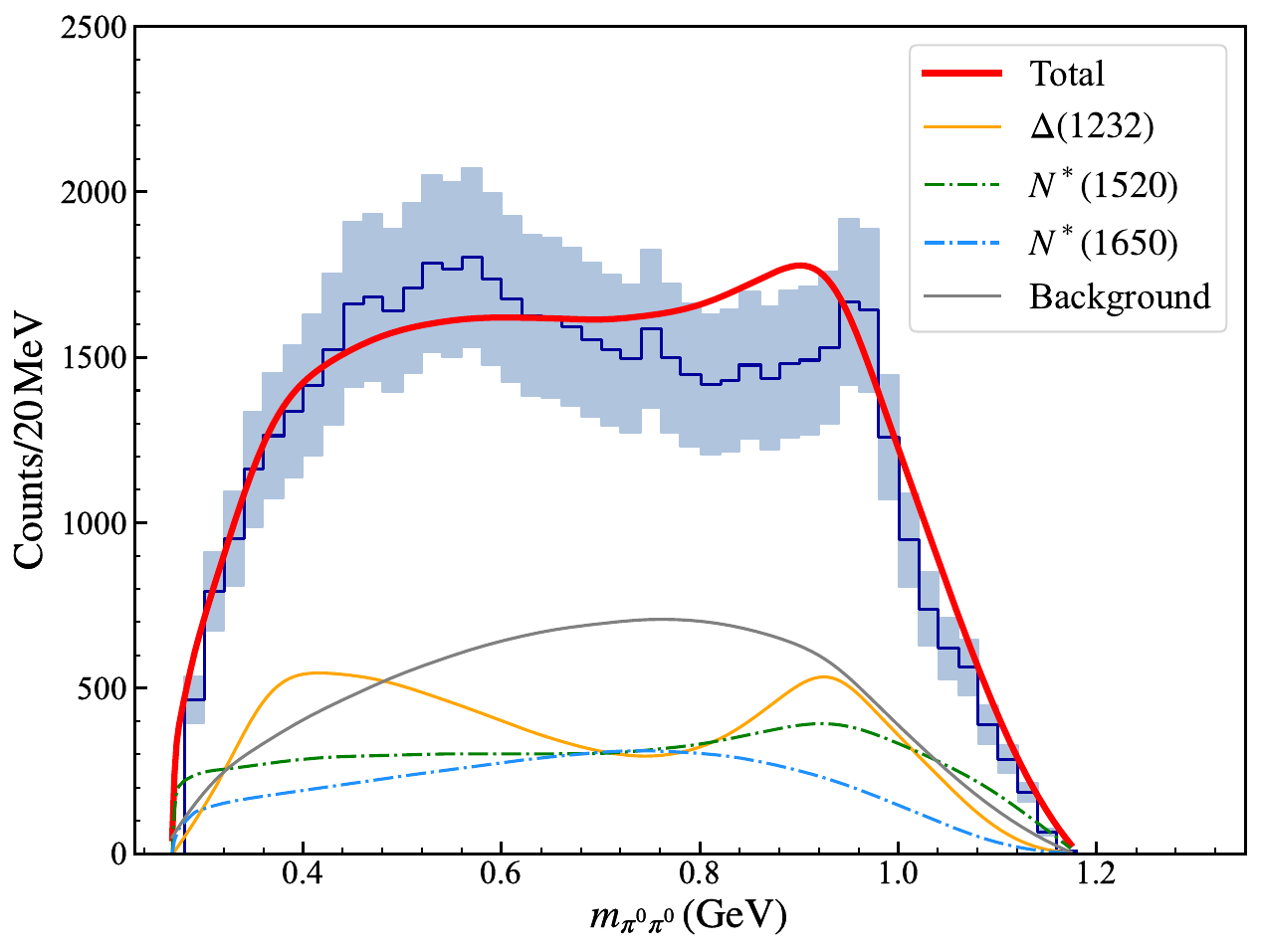}
    \caption{Invariant mass spectra for the ${\pi }^{0}{\pi }^{0}$ pair in the total energy range $ 1898< W < 2110 $~MeV  with experimental data taken from Ref.~\cite{LEPS2BGOegg:2023ssr}.}
    \label{fig:massc-1898-2110-pi0pi0-low}
\end{figure}

\begin{figure}[htbp]
    \centering
    \includegraphics[scale=0.37]{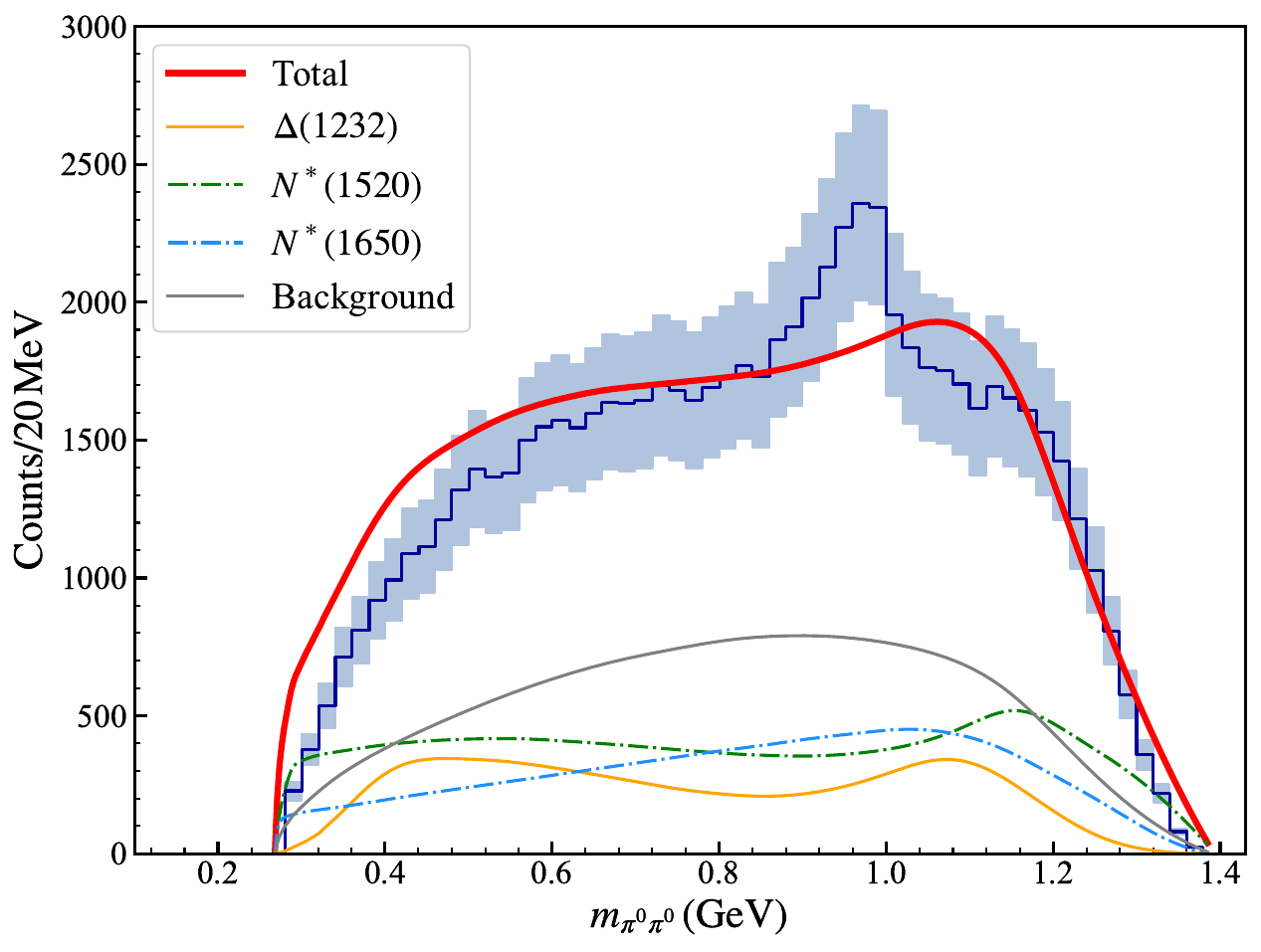}
\caption{As in Fig.~\ref{fig:massc-1898-2110-pi0pi0-low}, but for the invariant ${\pi }^{0}{\pi }^{0}$ mass distributions in the total energy range $ 2110< W < 2320 $~MeV. }
    \label{fig:massb-2110-2320-pi0pi0-high}
\end{figure}

Therefore, to achieve a more accurate comparison with experimental results, we introduced two overall factors $C_{\rm low}$ and $C_{\rm high}$ in the low-energy and high-energy regions of the ${\pi }^{0}{\pi }^{0}$ invariant mass spectrum as follows,
\begin{eqnarray}
&&0.73\int \frac{d\sigma }{dm_{\pi^0p}} dm_{\pi^0p}=\frac{1}{2} C_{\rm low}\int_{\rm low} \frac{d\sigma }{dm_{\pi^0\pi^0}} dm_{\pi^0\pi^0} \nonumber\\
&&+\frac{1}{2} C_{\rm high}\int_{\rm high} \frac{d\sigma }{dm_{\pi^0\pi^0}} dm_{\pi^0\pi^0},\nonumber\\
&&\frac{\frac{1}{2} C_{\rm low}\int_{\rm low} \frac{d\sigma }{dm_{\pi^0\pi^0}} dm_{\pi^0\pi^0}}{\frac{1}{2} C_{\rm high}\int_{\rm high} \frac{d\sigma }{dm_{\pi^0\pi^0}} dm_{\pi^0\pi^0}}=0.75.
\label{Theory-the area ratio of the higher and lower}
\end{eqnarray}
Then we could obtain he values $C_{\rm low} = 0.68$ and $C_{\rm high} = 0.77$, respectively. With these values,  in Fig.~\ref{fig:massc-1898-2110-pi0pi0-low} and Fig.~\ref{fig:massb-2110-2320-pi0pi0-high} we show the theoretical results of $d\sigma /d{{m}}_{{\pi }^{0}{\pi }^{0}} $ for the ${\pi }^{0}{\pi }^{0}$ invariant mass distributions at the lower ($ 1898< W< 2110 $~MeV) total energy region and higher one ($2110< W < 2320 $~MeV), comparing with the experimental measurements taken from Ref.~\cite{LEPS2BGOegg:2023ssr}. One can see that the theoretical calculations can roughly reproduce the experimental data except the bump structures for the $f_0(500)$ and $f_0(980)$ mesons, which were not included in present model. The contribution of $\Delta(1232)$ gives two bump structures in the invariant $\pi^0 \pi^0$ mass distributions.

As discussed in the Introduction part, the contribution from the $f_0(980)$ meson is important to the $\gamma p \to \pi^0 \pi^0 p$ reaction. However, the current data are limited, and we leave the study of $f_0(980)$ meson in future work when more experimental data are available.

\section{Summary}

In this work, the experimental data on the differential cross section of the $\gamma p\to {\pi }^{0}R\to{\pi }^{0}{\pi }^{0}p$ ($R\equiv\Delta (1232), {N}^{*}(1520)$, and ${N}^{*}(1650)$) reaction provided by the LEPS2/BGOegg Collaboration~\cite{LEPS2BGOegg:2023ssr} are analyzed by using the effective Lagrangian approach, where the tree level diagrams of $s$-channel nucleon pole and $t$-channel $\rho^0$ exchange are considered. We study the $\Delta(1232)$, $N^{*}(1520)$, and $N^{*}(1650)$ resonances decay into $\pi^{0}p$ through $P$-wave, $D$-wave, and $S$-wave, respectively. It is found that in the currently considered energy region, the contributions of the $\Delta (1232)$ in the $s$-channel and the $N^*(1520)$ and $N^*(1650)$ in the $t$-channel are dominated. Moreover, it is interesting to expect that future measurements on this reaction will offer a further test of the $\rho N {N}^{*}(1520)$ and $\rho N N^*(1650)$ couplings and help us to better understand the role of vector meson exchange in relevant reactions. 

Finally, regarding the contributions of $\Delta(1232)$, $N^{*}(1520)$, and $N^{*}(1650)$ resonances in the $\pi^0 p$ invariant mass spectrum, it is hoped that more theoretical works and experimental measurements can be incorporated, so as to further study the properties of the low-lying baryon excited states. Furthermore, more precise data about the $\gamma p \to \pi^0 \pi^0 p$ reaction can be also used to study the nature of the scalar mesons $f_0(500)$ and $f_0(980)$. 

\section*{Acknowledgements}

This work is partly supported by the National Key R\&D Program of China under Grant No. 2023YFA1606703, No. 2024YFE0105200, and the Natural Science Foundation of Henan under Grant No. 232300421140 and No. 222300420554. It is also supported by the National Natural Science Foundation of China under Grant Nos. 12435007, 12361141819, 12475086, and 12192263.

\normalem
\bibliographystyle{apsrev4-1.bst}
\bibliography{ref.bib}

\end{document}